\documentclass[reprint,aps,showkeys,preprintnumbers,nofootinbib,showpacs, floatfix, twocolumn, superscriptaddress]{revtex4-1}

\usepackage{graphicx}
\usepackage[dvipsnames]{xcolor}
\usepackage[shortlabels]{enumitem}
\usepackage{mathtools}
\usepackage{fancyhdr}
\usepackage{nicefrac}
\usepackage{mathrsfs}
\usepackage{subcaption}
\usepackage{quantum}
\usepackage{bbold}
\makeatletter

\usepackage{amsfonts,amssymb,amsmath}            
\usepackage{ifthen}
\usepackage{amsthm, thm-restate}
\usepackage{dsfont}
\usepackage{float}                       
\usepackage{cancel}

\PassOptionsToPackage{hyphens}{url}
\usepackage{hyperref}
\usepackage{xcolor}
\definecolor{mylinkcolor}{rgb}{0,0,0.7} 
\hypersetup{unicode=true,%
  bookmarksnumbered=false,bookmarksopen=false,bookmarksopenlevel=1, %
  breaklinks=true,pdfborder={0 0 0},colorlinks=true}%
\hypersetup{%
  anchorcolor=mylinkcolor,citecolor=mylinkcolor, %
  filecolor=mylinkcolor,linkcolor=mylinkcolor, %
  menucolor=mylinkcolor,runcolor=mylinkcolor, %
  urlcolor=mylinkcolor}

\DeclareRobustCommand\ground{\begin{tikzpicture}[scale=0.8]
		\draw [thick](-0.4,0)--(0.4,0);\draw [thick](-0.3,0.1)--(0.3,0.1);\draw [thick](-0.2,0.2)--(0.2,0.2);\draw [thick](-0.1,0.3)--(0.1,0.3);
	\end{tikzpicture}}

\usepackage[capitalise]{cleveref}

\usepackage{tikz}
\usepackage{soul}
\usepackage{hyperref}
\usetikzlibrary{arrows}
\usetikzlibrary{arrows.meta}
\usepackage{rotating, xcolor}
\usetikzlibrary{shapes}
\usetikzlibrary{hobby}
\usetikzlibrary{decorations.markings}
\usetikzlibrary{calc,intersections,through,backgrounds}
\usetikzlibrary{decorations.pathreplacing,angles,quotes}
\usetikzlibrary{arrows,decorations.markings}
\usetikzlibrary{arrows.meta}

\theoremstyle{plain}

\theoremstyle{definition}
\newtheorem{definition}{Definition}[section]

\usepackage[nodayofweek]{datetime}
\newpage

\begin{document}

\title{Events and their Localisation are Relative to a Lab}

\date{\today}
\author{V.Vilasini}
\affiliation{Universit\'{e} Grenoble Alpes, Inria, 38000 Grenoble, France}
\affiliation{Institute for Theoretical Physics, ETH Z{\"u}rich, 8093 Z{\"u}rich, Switzerland}
\author{Lin-Qing Chen}%
\affiliation{Institute for Quantum Optics and Quantum Information (IQOQI),Austrian Academy of Sciences}
\affiliation{Faculty of Physics, University of Vienna, Austria}
\affiliation{Institute for Theoretical Physics, ETH Z{\"u}rich, 8093 Z{\"u}rich, Switzerland}
\author{Liuhang Ye}
\affiliation{Institute for Theoretical Physics, ETH Z{\"u}rich, 8093 Z{\"u}rich, Switzerland}
\author{Renato Renner}
\affiliation{Institute for Theoretical Physics, ETH Z{\"u}rich, 8093 Z{\"u}rich, Switzerland}

\begin{abstract}
The notions of events and their localisation fundamentally differ between quantum theory and general relativity, reconciling them becomes even more important and challenging in the context of quantum gravity where a classical spacetime background can no longer be assumed. We therefore propose an operational approach drawing from quantum information, to define events and their localisation relative to a \emph{Lab}, which in particular includes a choice of physical degree of freedom (the reference) providing a generalised notion of “location". We define a property of the reference, relative measurability, that is sensitive to correlations between the Lab’s reference and objects of study. Applying this proposal to analyse the quantum switch (QS), a process widely associated with indefinite causal order, we uncover differences between classical and quantum spacetime realisations of QS, rooted in the relative measurability of the associated references and possibilities for agents' interventions. Our analysis also clarifies a longstanding debate on the interpretation of QS experiments, demonstrating how different conclusions stem from distinct assumptions on the Labs. This provides a foundation for a more unified view of events, localisation, and causality across quantum and relativistic domains.
\end{abstract}

\maketitle

\section{Introduction}
\label{Introduction}

Events and their localisation—what happens, where and when—are central to our understanding of the world, in both everyday life and science. They are closely tied to causality, which underpins a theory’s explanatory power, since events serve as the relata of causal connections. In quantum theory, they are defined operationally through measurements and interventions, often relative to observers or agents. In contrast, general relativity treats events as features of spacetime, independent of agents. This reveals a deep conceptual disparity between the two frameworks, one that must be bridged to consistently describe quantum experiments in spacetime and to understand the limits and possibilities for relativistic quantum information processing. The need for reconciliation is even more pressing and calls for a fundamental rethinking in quantum gravity, where spacetime is no longer a classical background.

We therefore aim to develop a general understanding of events and localisation without assuming a background spacetime, that can bridge quantum and relativistic perspectives. Instead of a specific theory, we begin with basic operational principles, by considering what information and degrees of freedom (d.o.f.) a laboratory can, in principle, access and act upon. A classic illustration is Maxwell’s demon \cite{Knott1911}, which appears to violate the second law of thermodynamics. Its resolution \cite{Landauer1961, Bennett1982} hinges on recognising that the demon and a macroscopic observer access different information and act on different d.o.f., and physical quantities like entropy are relative to this. This suggests that such relational, agent-dependent principles should also inform how events,  localisation, and causality are defined.

The following examples further illustrate key features relevant to our goal. A thermal system may be described either microscopically (e.g., by a demon acting on individual particles) or macroscopically (e.g., as the motion of a gas in a box). Cyclic causal structures capturing information-theoretic dependences between variables are widely used across classical statistical disciplines to model feedback processes \cite{Pearl_2009, Bongers_2021}, even when the underlying spacetime has an acyclic causal structure. Recognising that an agent could, in principle, intervene on each variable at different times allows these cycles to be resolved into acyclic causal sequences consistent with relativistic causality. Similarly, the causal link between the moon and tides is understood because, in principle, altering the moon’s mass or position would affect tidal patterns, even if such interventions are not feasible. These examples emphasize that accounting for in-principle possible operations enables reconciliation between information-theoretic and relativistic views, as also emphasized in \cite{VilasiniRennerPRA, VilasiniRennerPRL}. We thus seek a general definition of events, as motivated above, that can explain these features.

Our general goal naturally raises a key question: relative to what structure can events and their localisation be defined? To unambiguously achieve the goal, we propose that events should be defined with respect to three essential criteria: (a) a choice of level of detail at which a situation is analysed; (b) a choice of reference degrees of freedom used to condition operations on the systems of study (e.g., turning on a laser at 5pm, or sending a message when a quantum spin is “up”); and (c) a choice of allowed operations, that can include those possible in principle. These choices together define a Lab, an abstraction of a physicist in a laboratory who can observe and probe systems, and which captures implicit assumptions underlying physical reasoning across theories.

The concept of events is particularly essential for understanding the notion of indefinite causal order between quantum operations \cite{chiribella2013quantum,oreshkov2012quantum, araujo2015witnessing}. The quantum switch (QS) \cite{chiribella2013quantum} is a prominent example of this phenomenon. A longstanding question is whether (and how) proposed QS realisations in Minkowski spacetime (including several optical experiments \cite{Araujo_2014,Procopio_2015,Rubino2017,Goswami2018,Wei_2019,Guo_2020,Goswami_2020,Taddei_2021,Rubino_2021}) and in quantum gravity scenarios \cite{zych2019bell, Castro-Ruiz:2019nnl, Paunkovic2020,Moller_2021,Moller_2024} fundamentally differ in the properties of events, localisation, or causality. Recent work \cite{delaHamette:2022cka, Kabel:2024lzr,de2024event} suggests they do not. In contrast, applying our proposed definitions to analyze these physical realisations of QS in more detail reveals key distinctions within and across QS realisations (Tables~\ref{table:main}, \ref{table:appendix}). The interpretation of QS experiments has long been debated, reflecting divergent views of events and causality \cite{chiribella2013quantum,Procopio_2015,Portmann2017,Vilasini_mastersthesis, Oreshkov2019,Paunkovic2020,VilasiniRennerPRA,VilasiniRennerPRL,Ormrod_2023,Felce_2022,delaHamette:2022cka, Kabel:2024lzr}. We show how these distinct perspectives arise from differing assumptions about (a)–(c). This highlights that the identification, localisation and causal relations between events generally does depend on the choice of a Lab.

\section{More to the double-slit than meets the eye}
\label{sec: doubleslit}

We begin with the familiar double-slit experiment \cite{Young1804}
to intuitively illustrate our definitions and demonstrate how re-examining well-known scenarios  through the lens of criteria (a)–(c) can reveal new insights.  

A minimal quantum description would only include a unitary evolution from the particle emission to detection on the screen, providing no information on intermediate details, such as “the particle passed through one of two slits”. To recover the more standard quantum  description of the experiment, a number of choices need to be made. For example, we must describe the (sub)systems on which this unitary acts and identify a preferred basis that allows to interpret this unitary as describing a well-localised wall with slits and the particle passing through them in a superposition. These specify the level of detail (criterion (a)) typically assumed when describing the experiment.

{\bf In-principle possible interventions} The implicit structure underlying this level of detail is provided by an ``agent'' (such as us) capable, in principle, of making interventions to understand the experiment, including measuring the presence of the wall. These interventions inherently rely on reference d.o.f.\  (e.g., clocks and rods), thus invoking criteria (b) and (c). Consider, for example, a classical experimentalist, Claire, who can independently intervene at the two slits using physical spacetime references available to her\footnote{In finite-dimensional quantum theory, this operational ability to independently prepare, measure, or act on states at each location implies that the Hilbert space associated with the two slit positions takes a tensor product form for Claire.} e.g, she may place phase shifters at the spatial location of each slit. This enables the identification of two intermediate events, which causally influence the interference pattern (e.g., by shifting the fringes).

A recent work \cite{de2024event} proposed ``what an event is not", arguing that “$E_{\textnormal{left/right}}$: the particle passing through the left/right slit in the interference experiment” should not be regarded as two distinct events, as they cannot be distinguished without disturbing the ``phenomenon''. The two operational events that Claire can identify at the two slits, as we describe above, are distinct from the ``non-events'' $E_{\textnormal{left/right}}$, and therefore do not conflict with the proposal of \cite{de2024event} or quantum complementarity. However, this highlights that the proposal of \cite{de2024event} prompts further questions: if the interference pattern defines the phenomenon, does altering it by introducing phase shifts at each slit constitute a disturbance? More fundamentally, information-disturbance relations from quantum information imply that measurement inherently disturbs non-classical systems proportionally to information gained.
If (non-)events are defined by operational (in)distinguishability without disturbance \cite{de2024event}, does this mean that only occurrences known to happen with certainty count as events? Addressing such questions requires making explicit the choices (in particular of (a)-(c)) relative to which one defines events.

{\bf A more exotic double-slit experiment} Consider a hypothetical quantum experimentalist, Quinn, whose laboratory is initially prepared by Claire in a state entangled with the particle. The combined system, the particle and Quinn’s lab, traverses the double-slit in a superposition, specifying the initial condition for Quinn's lab\footnote{In principle, Quinn can have knowledge of this initial condition and how he is set-up within the experiment.}. Quinn, in contrast to Claire, can only intervene locally relative to his lab and is not localised relative to the background spacetime. Hence, unlike Claire, he cannot independently intervene at each slit due to constraints imposed by his quantum trajectory. Claire and Quinn represent distinct Labs, each equipped with different reference d.o.f.\  and sets of physically allowed operations.

A key distinction between Claire and Quinn lies in their spatiotemporal references: Claire’s measurement of her spatial reference (e.g., slit positions) or of time on her lab's clock does not affect the particle or its interference pattern in any way, while Quinn’s spatial measurement (e.g., relative distance to center of slits) would since this gives him which-way information. Hence, Claire’s spatial reference is called \emph{relatively measurable}, while Quinn’s is not (given his initial condition). Moreover, Claire's operations are \emph{localised} relative to her spatial reference, whereas Quinn’s operations lack such localisation due to his own spatial non-localisation (given by his initial conditions). Thus Quinn, unlike Claire, cannot identify two spatially localised operational events at the slits.

This sets the stage for our formal definitions and a key distinction we later make between quantum switch realisations.

\section{Generalised notions of events and localisation}
\label{sec: definitions}

Our proposal is based on the following operational assumptions: 

\begin{enumerate}
    \item It is possible to distinguish different degrees of freedom (d.o.f.), in particular: target d.o.f.\  which are the objects of study and reference d.o.f.\  which are used to condition operations performed on the target\footnote{It is important to note  a clear distinction: the notion of reference here is different from reference frames, and in particular from the recent developments of quantum reference frames \cite{Bartlett_2007, Loveridge_2018, Giacomini_2019, Vanrietvelde_2020, delaHamette:2021oex, Castroruiz_2023, Carette_2025}. For instance, a quantum spin can serve as a reference d.o.f.\  used to condition the operation when it appears as the control system of a CNOT gate — a role distinct from the situation when a spin serves as a quantum reference frame for other spins. }. A d.o.f.\  is mathematically associated with a well-defined space (e.g., a space of states or one specified by an algebra).

    \item It is possible to measure these d.o.f.\  and observe distinguishable outcomes. A measurement on a d.o.f.\  is  described by a set of operators acting on that d.o.f., with one operator corresponding to each possible outcome.
    
\end{enumerate}

\begin{definition}[Lab and relative events]
\label{def:lab}
    A \emph{generalised lab} or Lab $A$ is specified by:
    \begin{itemize}
        \item A \emph{reference information d.o.f.} $R_A$ (which we will call reference d.o.f. for brevity), along with a measurement $\mathcal{M}_{R_A}:=\{\Pi_\lambda\}_{\lambda\in \Lambda}$ that can be performed on it, whose physical outcome labels $\mathcal{P}_A:=\Lambda$ (referred to as the \emph{reference information set}) decompose the space of $R_A$ into distinct subspaces. We refer to such a measurement as a $\mathcal{P}_A$-level measurement.
        \item A \emph{target d.o.f} $T_A$
        \item A \emph{set of operations} $\mathcal{O}_A$ on $R_A$ and $T_A$ that consist of operations of the following form: conditioned on the $\lambda$ subspace of $R_A$, a corresponding operation $O_A^\lambda$ acts on $T_A$.
    \end{itemize}

Then, each $O_A\in \mathcal{O}_A$ is said to be an event relative to Lab $A$, or simply a \emph{relative event} (if the Lab is clear from context).
\end{definition}

An example of conditional operations described above are controlled operations in quantum information theory which include both coherent and incoherent quantum control. \cref{appendix: existing_notions} provides further details on the form of $O_A$. It will be convenient (especially in \cref{table:main}) to represent the reference information set $\mathcal{P}_A=\Lambda$ as $\mathcal{P}_A=\{\lambda\}$ when needing to highlight the individual values $\lambda$.
Finally, and most importantly, a specification of a Lab incorporates a choice of criteria (a)-(c). The description of $R_A$ and $T_A$ is based on the choice of a level of detail (a), and which d.o.f is used as reference (b), and the choice of allowed operations (c) is also incorporated in the Lab.

We now define the concept of \emph{relative measurability} of the reference and \emph{localisation} of events relative to a Lab. These concepts were motivated in the example of \cref{sec: doubleslit} (and will appear in the examples of \cref{sec: QSanalysis}). In particular, the spatial reference of Claire in \cref{sec: doubleslit} is relatively measurable while that of Quinn, who travels in superposition along with the target photon, is not.  This requires an additional ingredient: the initial conditions (e.g., state preparation) $\mathcal{I}_A$ of the Lab $A$, defined on the reference $R_A$ and target $T_A$ of $A$ along with an additional \emph{environment d.o.f} $E_A$ which is external to the Lab $A$ but may be correlated with $R_A$ and $T_A$.\footnote{The initial conditions $\mathcal{I}_A$ of Lab $A$ may be prepared by another Lab (e.g., be specified among the operations of that Lab). However the information may be known e.g., stored in the memory of an agent therein (see \cref{sec: outlook}), to the Lab $A$. }

\begin{definition}[Relative measurability of the reference d.o.f]
\label{def: measurability}
    Given a Lab $A$ and initial conditions $\mathcal{I}_A$, we say that the Lab's reference $R_A$ is \emph{relatively measurable at an event $O_A\in \mathcal{O}_A$} if the condition expressed in the following diagram holds, where \ground\ denotes discarding a d.o.f.
\begin{center}
      \includegraphics[scale=1.0]{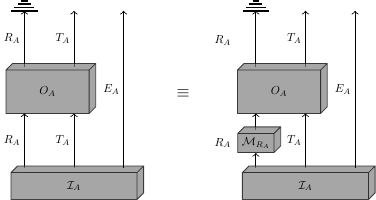}  
\end{center}

    If this holds for all $O_A\in \mathcal{O}_A$, we simply say that $R_A$ is \emph{relatively measurable}. 
\end{definition}

The above condition captures that for the given $\mathcal{I}_A$, applying $O_A$ has the same effect on $T_A$ and $E_A$ as performing a $\mathcal{P}_A$-level measurement on $R_A$ and then applying $O_A$.

\begin{definition}[Localisation of a Lab's relative events]
\label{definition: localisation}
Given a Lab $A$ and initial conditions $\mathcal{I}_A$, a relative event
    $O_A\in\mathcal{O}_A$ is said to be a \emph{$\lambda$-localised relative event} if the condition expressed in the following diagram holds
    \begin{center}
        \includegraphics[scale=1.0]{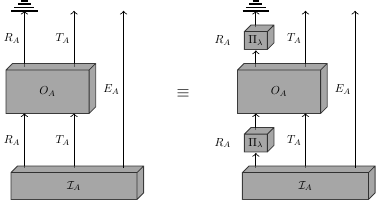}
    \end{center}

   If there exists no $\lambda\in \mathcal{P}_A$ for which this is the case,  we say that $O_A$ is a \emph{non-localised relative event}.       
   
\end{definition}

\begin{figure*}[ht!]
    \hspace*{-1.5cm}
    \subfloat[QS$_{CT}$]{\includegraphics[scale=0.26]{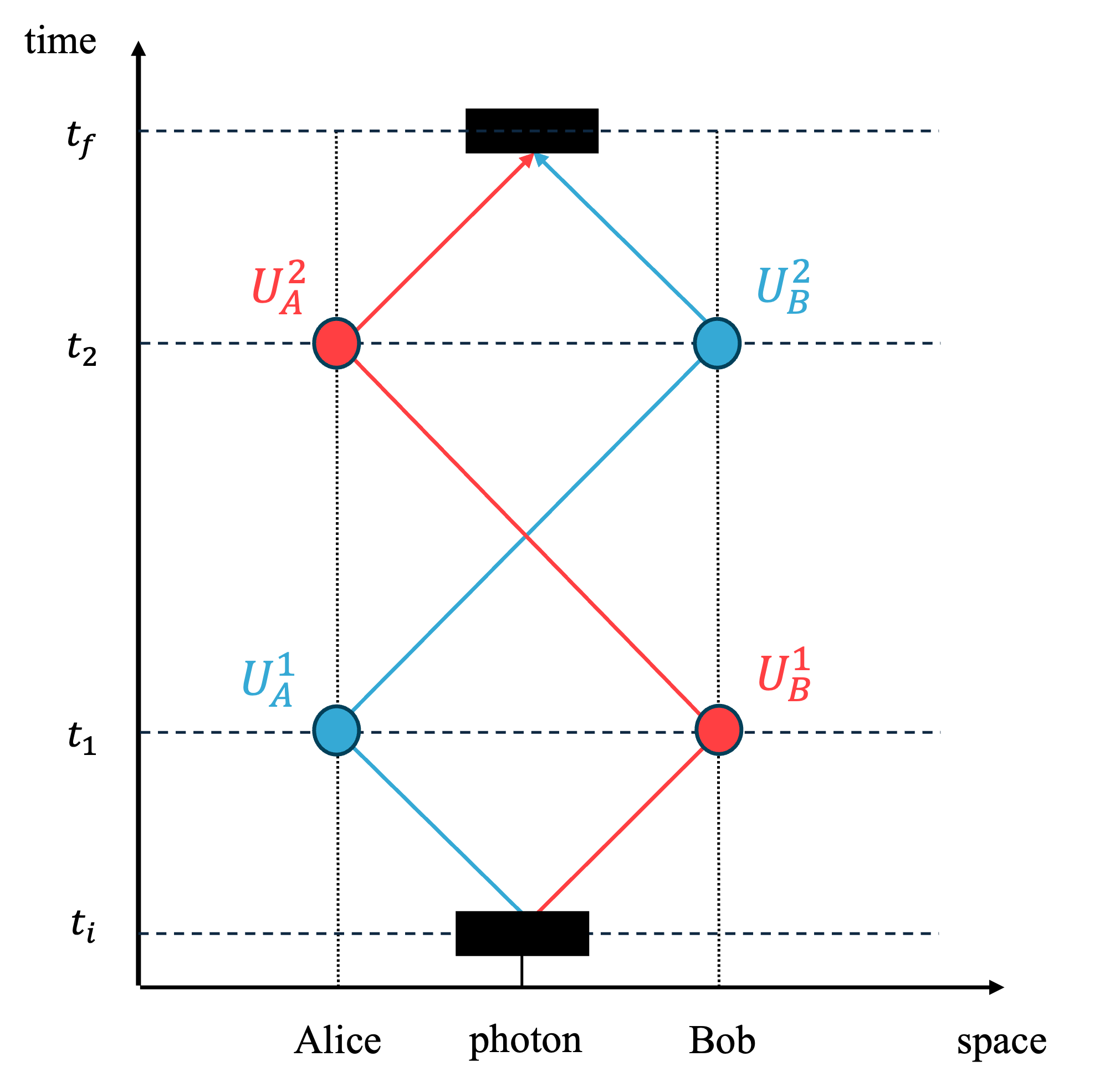}}\subfloat[QS$_{QT}$]{\includegraphics[scale=0.31]{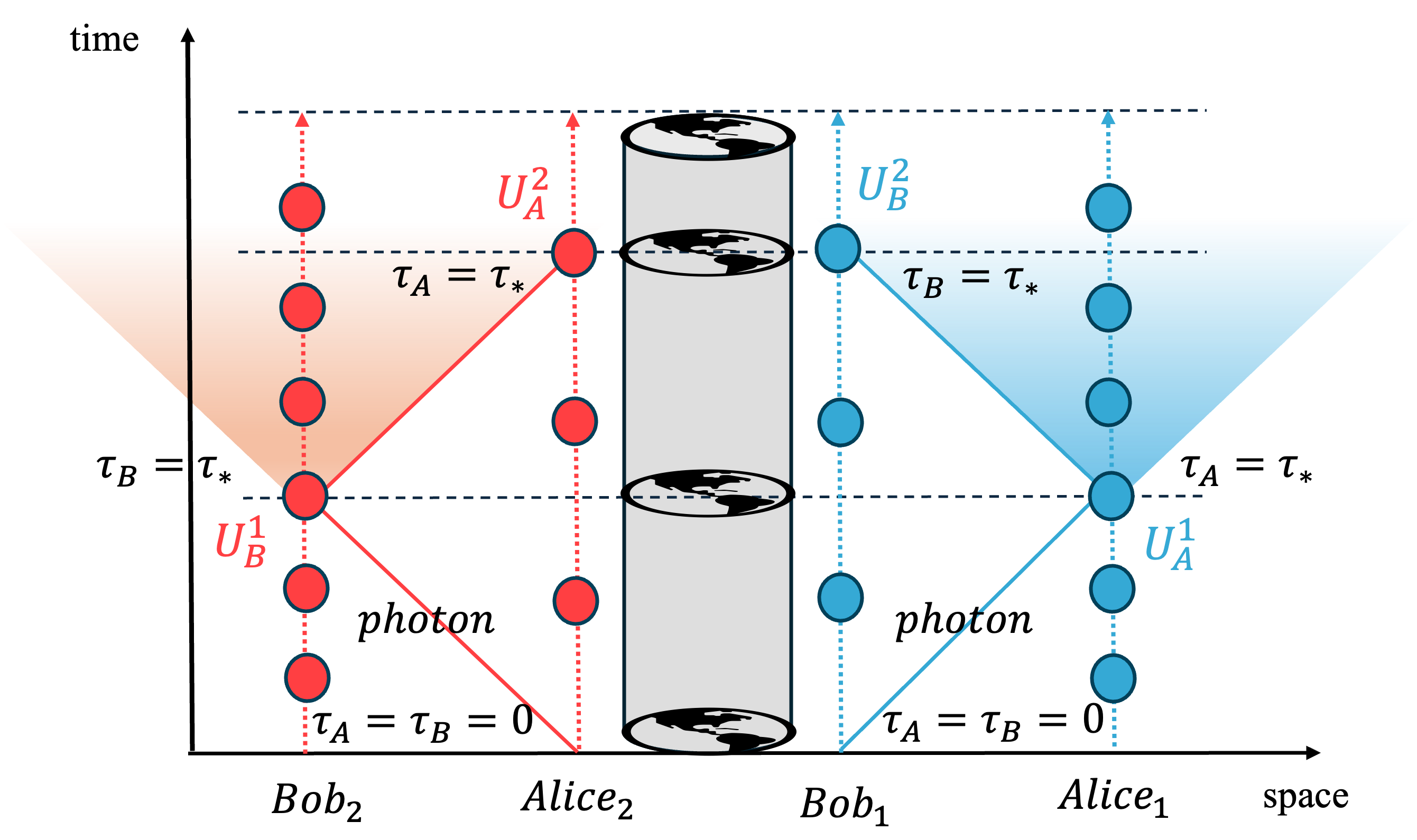}}
    \subfloat[QS$_G$]{\includegraphics[scale=0.31]{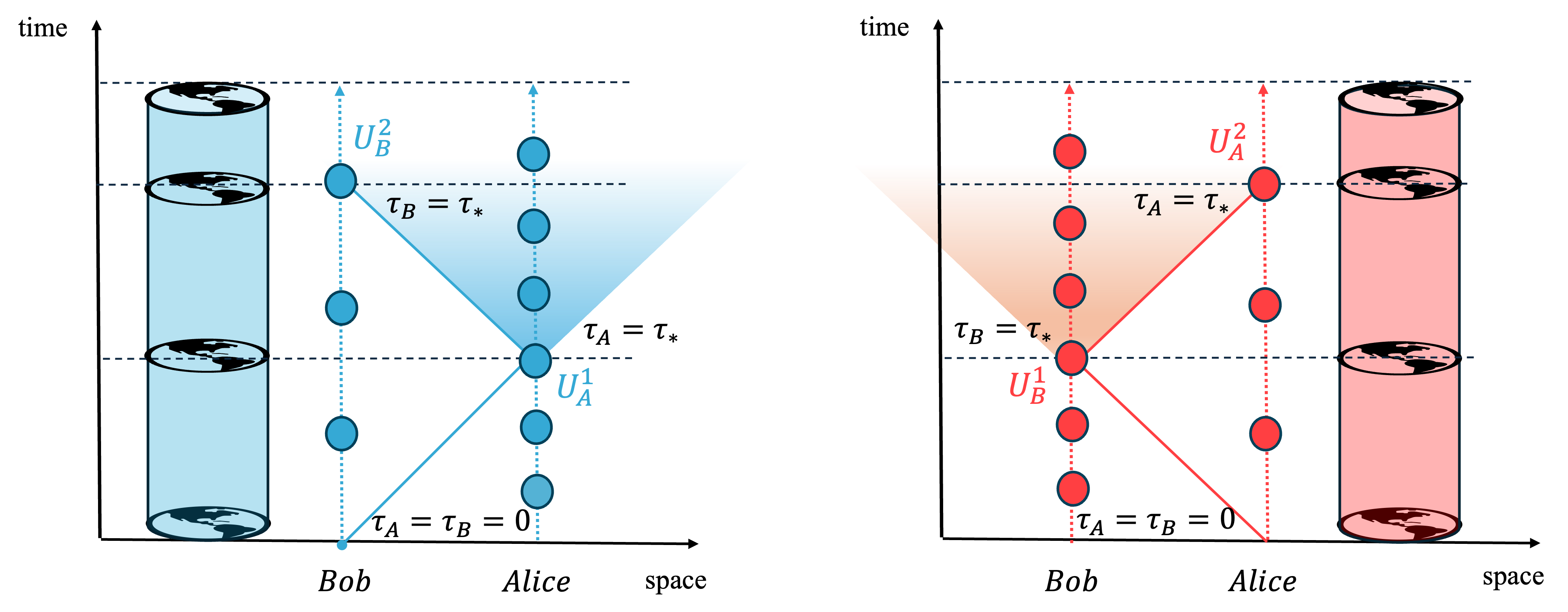}}
    \caption{
    Three quantum switch (QS) protocols. (a) QS$_{CT}$: Alice and Bob each follow a classical trajectory (black dotted lines) in the classical background spacetime, the target system is in a superposition of trajectories (red and blue solid lines). This corresponds to the spacetime diagram of a typical QS$_O$ experiment \cite{Procopio_2015}.
    (b) QS$_{QT}$: Alice and Bob take superpositions of trajectories (red and blue dotted lines) in the classical background spacetime, so does the target (red and blue solid lines). By moving all the Alice/Bob trajectories to one side of the mass, preserving relative distances, we can have an equivalent situation where the target has a classical but Alice/Bob a quantum trajectory. The circles denote equal \emph{proper time} intervals of the corresponding trajectories. (c) QS$_G$ \cite{zych2019bell}: a superposition of spacetime geometries drawn separately in the left/right diagrams. Alice and Bob follow different trajectories in the two branches (red and blue dotted lines), so does the target (red and blue solid lines). The gravitational field configurations between (b) and (c) can be related by a quantum reference frame transformation \cite{delaHamette:2021iwx}.}
    \label{fig:examples_QS}
\end{figure*}

The diagrammatic representation used above is common in quantum information theory. This has a precise formal meaning in a general class of operational theories including quantum theory (see for example \cite{Coecke_Kissinger_2017, Chiribella2010}). In quantum information protocols, the environment d.o.f. $E_A$ corresponds to d.o.f. in the parts of the protocol external to a Lab $A$, making the notions of relative measurability and localisation inherently dependent not only on the Lab but also on the context of the surrounding protocol.

In \cref{appendix: existing_notions}, we further discuss how these definitions can recover as special cases, existing notions of events and localisation such as crossing of world-lines in general relativity and sub-system localisation in quantum theory.

\section{Quantum switches in classical vs quantum spacetimes and labs}
\label{sec: QSanalysis}

The quantum switch (QS) \cite{chiribella2013quantum} is an abstract information-theoretic protocol associated with \emph{indefinite causal order} (ICO), where two operations—e.g., unitaries $U_A$ (for Alice) and $U_B$ (for Bob)—are applied to a target system in a quantum superposition of different orders, coherently controlled by a control system. A formal review of QS is provided in Appendix \ref{QS review}. Several optical table-top experiments \cite{Araujo_2014,Procopio_2015, Rubino2017, Goswami2018, Wei_2019, Guo_2020, Goswami_2020, Taddei_2021, Rubino_2021} in Minkowski spacetime (optical quantum switch, QS$_O$) claim to physically implement the ICO of QS, while thought experiments \cite{zych2019bell, Paunkovic2020,Moller_2021, Moller_2024} explore its realisation in quantum-gravitational settings (gravitational quantum switch, QS$_G$). However, the interpretation of QS$_O$ experiments and their connection to QS$_G$ remain subjects of ongoing debate \cite{chiribella2013quantum,Procopio_2015,Portmann2017,Vilasini_mastersthesis, Oreshkov2019,Paunkovic2020,VilasiniRennerPRA,VilasiniRennerPRL,Ormrod_2023,Felce_2022,Kabel:2024lzr} without clear consensus.

To further clarify these aspects, we analyze three versions of the QS protocol (illustrated in \cref{fig:examples_QS}). The first two occur in classical spacetime, while the third corresponds to QS$_G$ (as proposed in \cite{zych2019bell}), a thought experiment in a regime of quantum spacetime (superposition of semi-classical geometries). The classical cases differ in whether Alice and Bob follow classical trajectories or quantum superpositions of trajectories relative to the classical background spacetime, analogous to Claire and Quinn in \cref{sec: doubleslit}. We refer to these as QS$_{CT}$ and QS$_{QT}$, where $CT$ and $QT$ stand for classical and quantum trajectories respectively. QS$_{CT}$ corresponds to a particular type of an optical quantum switch QS$_O$ experiment, where we make explicit that it is being analysed by a classical lab.

Applying our proposed definitions of relative events and localisation, by considering Labs associated with different choices of criteria (a)-(c) from \cref{Introduction}, leads to fundamental differences in the localisation and measurability properties between QS$_{CT}$ vs QS$_{QT}$ and QS$_G$. This is summarized in \cref{table:main}. In this analysis, we consider choices of Lab $A$ associated with the agent Alice of the QS. The initial conditions are specified by the QS protocol in consideration, and we take the environment $E_A$ to be given by the control qubit of the QS (which is external to the agents' Labs).

A detailed formal analysis of these conclusions, the underlying assumptions, and their link to Hilbert space structure is provided in \cref{appendix: analysis}.

{\bf Distinguishing different QS realisations} Consider QS$_{CT}$, here Alice’s spatial $x$ and time $t$ references are relatively measurable by \cref{def: measurability} (as with Claire in \cref{sec: doubleslit}). For instance, this holds for $t$ because she can measure her lab's clock to determine $t$ without disturbing the QS experiment. However Alice’s operation extends over time—she receives a photon at $t_1$ or $t_2 > t_1$ (and a vacuum, denoting absence of the photon at the complementary time) in a coherent superposition, depending on the QS’s control. Applying \cref{definition: localisation} confirms that Alice's operation is non-localised in time but spatially localised at $x_A$. Furthermore, Alice can independently intervene at $t_1$ and at $t_2$, corresponding to more fine-grained operations \cite{VilasiniRennerPRA,VilasiniRennerPRL} (see also \cref{fig:QS_fine}), called A$_1$, A$_2$’s operation in \cref{table:main}, and indeed these operations are time localised by \cref{definition: localisation}.

In QS$_{QT}$, Alice’s operation is localised in her proper time $\tau_A$, which remains a relatively measurable reference. However, her spatial reference $x$ (relative distance to the mass) and the background time $t$ are no longer measurable, analogous to Quinn in \cref{sec: doubleslit}. The background time $t$ corresponds to the proper time of an external, spacetime-localised agent (call her Claire) whose lab encompasses the experimental setup. Since $t$ differs between the two branches when Alice interacts with the target as seen in \cref{fig:examples_QS}(b), Alice measuring $t$ during her target interaction would collapse her trajectory and the QS control, making this a relatively non-measurable reference. Moreover, this also implies that Alice’s operation is not localised in $t$. The analysis for Alice’s perspectives in QS$_G$ and QS$_{QT}$ are analogous, noting that her acceleration $a$ directly corresponds to her relative distance from the mass. Note that we can have different choices of Labs for the same agent, e.g., Alice, depending on the choice of physical reference used for localisation (criterion (b)) or whether fine-grained operations such as A$_1$ and A$_2$ are taken into account (criteria (a) and (c)).

Moreover, when QS$_{QT}$ is analysed from the Lab of the external experimenter Claire (who has a classical trajectory), the conclusions align with those of Alice in QS$_{CT}$ when using clock's time $t$, reflecting the time coordinate, as reference (lines 1 and 2 of \cref{table:main}).

{\bf Reconciling different perspectives}
Both QS$_{CT}$ and QS$_{QT}$ exhibit non-localisation in $t$, consistent with prior works linking indefinite causal order to absence of temporal localisation \cite{Oreshkov2019,zych2019bell,Giacomini_2019,VilasiniRennerPRA,VilasiniRennerPRL, kabel2024}. However, our analysis shows that the physical meaning of $t$ and its non-localisation differs for Alice in QS$_{CT}$ (or Claire in QS$_{QT}$) versus Alice in QS$_{QT}$, due to differences in measurability. Alice in QS$_{CT}$ and Claire in QS$_{QT}$ are localised agents in a background spacetime. These situations fall within a prior framework \cite{VilasiniRennerPRL, VilasiniRennerPRA} that showed that any physical protocol in a classical spacetime (when analyzed for agents' localised therein, who have access to spacetime references e.g., clocks) admits a physical fine-grained description with a definite, acyclic causal order consistent with relativity (see \cref{fig:QS_fine} for the fine-grained QS model). However, QS$_{QT}$ and QS$_G$, when analysed from the corresponding Alice’s (and Bob’s) perspective, fall outside the framework of \cite{VilasiniRennerPRL, VilasiniRennerPRA}, and no longer admit a similar physically meaningful fine-grained description in terms of a definite acyclic order for those quantum agents (even when this is the case for external experimenters like Claire). The conclusions regarding causal order in these cases aligns with those of \cite{Oreshkov2019,delaHamette:2022cka, Kabel:2024lzr}.

The proposed approach reconciles the differing conclusions in the debate over indefinite causal order in QS experiments -- they stem from different assumptions about the Lab (specifically, its references and allowed operations) in the analysis (see also \cref{appendix: analysis}).

\begin{table*}[ht!]
    \centering
    \begin{tabular}{|c|c|c|c|c|}
        \hline
      Protocols ($\mathcal{I}_A$)  & $\mathcal{P}_A$ of the Lab  & $O_A$ (relative event) & Rel. measurability  of $R_A$ & Localisation of $O_A$   \\ 
        \hline 
QS$_{CT}$ & $\{t\}$ & Alice's (A's) operation & Yes & non-localised\\
QS$_{CT}$ & $\{t\}$ & A$_{1}$, A$_2$'s operation& Yes & $t_{1/2}$-localised \\
QS$_{CT}$ & $\{x\}$ & A, A$_1$, A$_2$ operation& Yes & $x_A$-localised \\
QS$_{QT}$ & $\{(x,t)\}$ & A's operation & No & non-localised\\ 
   QS$_{QT}$,   QS$_G$  & $\{a\}$ & A's operation& No& non-localised\\ 
   QS$_{QT}$, QS$_G$ & $\{\tau\}$ & A's operation & Yes & $\tau_*$-localised\\ 
    All QS & $|\mathcal{P}_A|=1$ & A's operation& Yes & localised \\
        \hline
    \end{tabular}
    \caption{Summary of analysis of \cref{sec: QSanalysis}, for the QS protocols in \cref{fig:examples_QS}. All rows analyse Alice's perspective, the details of the protocol provide the initial conditions $\mathcal{I}_A$ of the Lab that incorporates the reference and target d.o.f., the environment d.o.f (here, the control of the QS), and the allowed operations of this perspective. The references $\tau$, $a$ denote proper time and acceleration of the Alice in consideration, and $x$, $t$ the spatial and temporal coordinates in classical spacetime protocols. In the last column, $t_{1/2}$, $x_A$ and $\tau_*$ denote specific values in the reference information set as specified by the protocols in \cref{fig:examples_QS}. In the last line, $|\mathcal{P}_A|=1$ denotes that the reference set $\mathcal{P}_A$ only has one element. For example, 
    this can be seen as just having the target label $T_A$  and any operation on $T_A$ is thus $T_A$-localised (consistent with sub-system localisation, see \cref{appendix: existing_notions}).}
    \label{table:main}
    \end{table*}

\section{Conclusions and broader outlook}
\label{sec: outlook}

We have proposed a general definition of events and their localisation without assuming a background spacetime or a particular theory, such as classical or quantum, but instead starting from basic operational assumptions. This enables the approach to be instantiated and further developed within broader formalisms, such as generalised, operational, or categorical probabilistic theories \cite{hardy2001quantum,Chiribella2010,dAriano2017,Coecke_Kissinger_2017}, and across different quantum and relativistic frameworks. Our  analyses of quantum switch (QS) experiments in classical and quantum spacetimes (\cref{table:main}, \cref{table:appendix}) leads to the conclusion: even for a fixed experiment, the appropriate theoretical description and the conclusions about events, localisation, and causal structure depend on the underlying Lab, defined by choices of (a) level of detail, (b) reference degrees of freedom, and (c) allowed operations.

A natural question following our analysis in \cref{table:main} is whether the Lab of a distant classical observer (Claire$_G$) in the gravitational QS (QS$_G$) has the same operational status as the classical Labs of Alice and Bob in QS$_{CT}$ (or the optical QS). In QS$_{CT}$  (\cref{fig:QS_fine}), if the Labs have access to  clocks with sufficient resolution, and the possibility of independent interventions using such  clocks, then it allows an unraveling into a definite and acyclic causal order (see \cref{fig:QS_fine}, \cref{appendix: analysis}). Can Claire$_G$ achieve the same in QS$_G$? The answer is non-trivial, it depends on whether she can operationally utilize her asymptotical references to perform independent interventions within the quantum gravity zone formed by the superposed masses. We address this in a follow-up work.

The concept of relative measurability introduced here captures a necessary property for the reference to act as ``classical location labels'' relative to the target under study. It is motivated by a key feature of non-classical systems, that measurements can disturb the system and its correlations with the environment. In the context of the quantum switch, we demonstrated that relative measurability of the reference d.o.f.\  allows to distinguish between optical and gravitational versions of the QS. However, this distinction arises not from gravity but from quantum correlations between spatiotemporal references and the phenomena under study. It would be interesting to precisely link relative measurability to concrete non-classical signatures in future work.

 Our work has focused on the perspective of a single Lab and we have not formally defined causality in our approach. However, our analysis of the quantum switch scenarios, showed how this approach can provide new insights into causality, and its relational nature, in such examples. An key next step is to explore how different Labs’ perspectives and predictions can be compared and reconciled in general protocols, which is  crucial for understanding causality and the emergence of objectivity within a relational, perspective-dependent framework. Our approach hints that when considering physical experiments from the view of laboratories that follow classical trajectories in the same classical spacetime, the spatiotemporal references of the associated Labs can be aligned to agree with the partial order of the spacetime’s light cone. In particular, this restriction (along with assuming quantum theory) would recover a previous framework \cite{VilasiniRennerPRL, VilasiniRennerPRA} applicable to quantum protocols respecting relativistic causality in classical spacetimes.

Beyond classical spacetimes,  comparing the properties reference sets of different Labs and deriving conditions for common agreement on events and causal structure between different Labs' perspectives, could yield new insights. This could be explored in a theory-independent axiomatic manner or theory/model specific way. In particular, based on the present approach, it would be interesting to develop an operational notion of quantum coordinate transformations that can apply beyond the regime of superposition of semiclassical spacetimes, by formulating them as transformations of physical reference information sets of different Labs. 
The current proposal provides a distinct notion of perspectives (relative to a Lab) and conclusions on events, localisation and causality as compared to analysis based on quantum reference frames  transformations \cite{Castro-Ruiz:2019nnl, delaHamette:2022cka, Kabel:2024lzr}, combining them could provide a more unified perspective.

Our work also highlights a contrast in the notion of events, localisation and causality between “classical” and “quantum” Labs of Claire and Quinn (\cref{sec: doubleslit}, \cref{sec: QSanalysis}). This raises deeper open questions, regarding the role of the associated agents: How does including (or excluding) a reasoning agent within Quinn’s Lab (which is in a superposition of trajectories), alongside the devices implementing operations in $\mathcal{O}_A$, affect events, localisation and causal structure? How do the operational properties of relative measurability and localisation (\cref{def: measurability}, \cref{definition: localisation}) depend on the knowledge and Heisenberg cuts of different agents? These questions naturally point to connections with another major research direction in quantum foundations: Wigner’s Friend \cite{Wigner1967} scenarios, which extend unitary quantum theory to include agents and their laboratories. Several no-go theorems in this field (e.g., \cite{brukner2015quantummeasurementproblem, Frauchiger2018, Brukner_2018, Bong2020}) reveal challenges for maintaining an absolute, observer-independent notion of events while remaining consistent with quantum predictions (and even those of a more general class of operational theories \cite{Vilasini_2019,Ormrod_2023measurement}). Recent works have also further elucidated their connections to other problems regarding causality and spacetime structure \cite{Cavalcanti2021,pienaar2024,Hausmann2025}. A key issue is the ambiguity of the Heisenberg cut: a classical outcome (inside the cut) observed by Quinn may appear in a quantum superposition (outside the cut) from Claire’s perspective.

In light of such challenges, relational frameworks for quantum measurements and consistent multi-agent reasoning across different Heisenberg cuts have been developed \cite{VilasiniWoods_2024,Ormrod_2024}. A promising next step is to extend our initial proposal and link it to relational frameworks for Wigner’s Friends, by explicitly modeling each Lab’s agent—specifically, their memory storing measurement outcomes—as a physical system interacting with $R_A$ and $T_A$. This would allow to formally address the afore-mentioned questions and pave a path toward a more unified treatment of relative events and causality, between the diverse communities and fields.

\medskip
\newpage
{\it Acknowledgements} We are grateful to Augustin Vanrietvelde for his detailed feedback on an early version of this manuscript that motivated significant improvements. We thank Viktoria Kabel for sharing her thoughts on the initial draft, and we thank her, \v{C}aslav Brukner,  and Anne-Catherine de la Hammette for insightful discussions. VV's acknowledges support from a government grant managed by the Agence Nationale de la Recherche under the Plan France 2030 with the reference ANR-22-PETQ-0007. LQC is supported by FWF ESPRIT fellowship with grant DOI: 10.55776/ESP390, and from the
ID\# 62312 grant from the John Templeton Foundation, as part of the ``The Quantum Information Structure of Spacetime, Second Phase (QISS 2)" Project. The opinions expressed in this publication are those
of the authors and do not necessarily reflect the views of the John Templeton Foundation. This work contributes to the COST Action CA23130 ``Bridging high and low energies in search of quantum gravity
(BridgeQG)" and COST Action CA23115 ``Relativistic Quantum Information (RQI)". We also acknowledge the support from  from the ETH Zurich Quantum Center and NCCR SwissMAP.

\newpage
\appendix

\section{Recovering existing notions and further details}
\label{appendix: existing_notions}

{\bf Recovering existing notions}
Here we discuss how the general definitions of events and localisation we have proposed in \cref{sec: definitions} can recover well known notions existing in different areas of physics, under specific choices of Labs.

In general relativity (GR), a common operational notion of events is given by intersections of worldlines. To describe such event in our definition, we need to make explicit the physical references $R_A$ one uses to define a coordinate system\footnote{Note that while physical reference frames can always be used to define coordinate systems, the converse is not true—not every coordinate system can be realised by physical reference frame.}. The observable outcome  of the references corresponds to an element in the reference information set $\mathcal{P}_A=\{(t,\vec{x})\}$. 
The systems (worldlines) form the target $T_A$; the intersection at $(t,\vec{x})$ can always be viewed as an operation $O_A$ conditioned on  $\mathcal{P}_A$ and it is a  $(t,\vec{x})$-localised event. 
By suitably choosing the reference set, one can also model localisation relative to time intervals or spacetime regions. 
Similarly, standard notions of events linked to particle scattering or quantum measurements in spacetime, can be recovered  as quantum operations controlled on the physical spacetime reference.

In quantum information, subsystem localisation is a widely used notion of localisation that does not refer to spacetime. For instance, in a Bell experiment, Alice and Bob apply operations ``locally’’ on subsystems $S_1$ and $S_2$ of an entangled pair. This can be captured by including the system labels in the reference information set $\mathcal{P}_A:=\{S_i\}_{i\in \{1,2,...\}}$, taking the target to be the set of systems $T_A:=\{S_i\}_{i\in \{1,2,...\}}$, and the operation set $\mathcal{O}_A$ to be such that when conditioned on the reference subspace associated to the label $S_i$, the operation on $T_A$ acts non-trivially (not the identity) only on the subsystem $S_i$. Then Alice and Bob's operations are $S_{1/2}$-localised relative events respectively.\footnote{When the protocol occurs in a background spacetime, this is also consistent with the spatio-temporal view where one may regard Alice and Bob's labs to be in different spatial locations $x_A$ and $x_B$ with the systems $S_1$ and $S_2$ sent to the two locations. Then including the spatial coordinate in $\mathcal{P}_A$, our definition would confirm that the operations are $x_{A/B}$-localised.}

{\bf Further details on a Lab's set of operations}
The main building blocks for defining the set of operations $\mathcal{O}_A$ of a Lab $A$ (\cref{def:lab}) are conditioned operations: conditioned on the $\lambda$ subspace of $R_A$ a corresponding operation $O_A^\lambda$ acts on $T_A$. These represent operations an agent in the Lab can perform or analyse on the systems of interest, using a generalised notion of “location” specified by the reference. However, the reference $R_A$ can also undergo its own physical dynamics—such as the natural evolution of a clock—that are not directly controlled by the agent. 

To allow both agent-controlled interventions and natural dynamics internal to the Lab, we do not require every $O_A \in \mathcal{O}_A$ to be described solely as a conditioned operation. This enables to naturally include time-extended events that may be composed of multiple finer-grained events. For instance, Alice may perform conditioned operations on a target at times $t_1$ and $t_2$, relative to her clock (reference), while the clock evolves from $t_1$ to $t_2$ in between. We can model this in two ways depending on the choice of Labs: as a single Lab of Alice with a time non-localised event, or as two Labs—Alice$_1$ and Alice$_2$—each associated with a time-localised event. This distinction, relevant in the fine-grained analysis of the quantum switch in \cref{sec: QSanalysis}, affects conclusions about causal relations between events. In more common life scenarios, such a description allows consider a conference as a single time-extended event, or a sequence of multiple events such as individual talks, depending on the context of interest.

\section{The quantum switch and its process matrix}
\label{QS review}

\begin{figure}[H]
	\centering
	\includegraphics[scale=0.6]{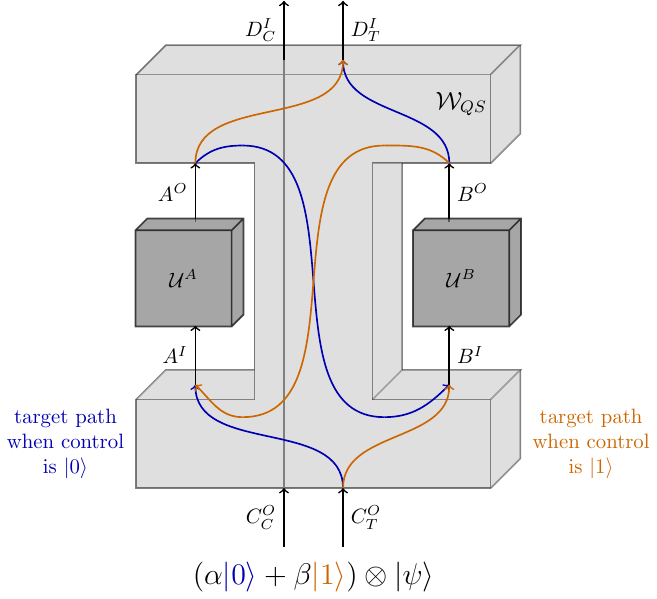}
	\captionsetup{font={small}}
	\caption{\textbf{Quantum switch as a process matrix:} $\hat{W}_{QS}$ represents the process matrix (or equivalently supermap) of the quantum switch. The target undergoes the blue path when the control is $\ket0$ and undergoes the orange path when the control is $\ket1$. Figure adapted from \cite{VilasiniRennerPRA}.} 
\label{fig: QS_PM}
\end{figure}

The quantum switch (QS) \cite{chiribella2013quantum} is a higher-order transformation, which given two operations $U_A$ and $U_B$, associated to agents Alice and Bob, applies them in quantum superposition of the two orders on a target system, depending coherently on a control qubit. Specifically, denoting the control Hilbert space as $\mathcal{H}_C$ and target Hilbert space as $\mathcal{H}_T$, the quantum switch implements the following transformation for any given unitaries $U_A$ and $U_B$, where $\ket{\Psi}$ is an arbitrary state of the target (see also \cref{fig: QS_PM}). We take the target to be a $d$ dimensional quantum system, a qudit.

\begin{equation}
\label{eq: QS_action} 
\begin{aligned}
    &\quad(\alpha \ket0+\beta \ket1)_C \otimes \ket \Psi_T \\ &\mapsto \alpha \ket0_C \otimes U_B U_A \ket \Psi_T + \beta \ket1_C \otimes U_A U_B \ket \Psi_T.
\end{aligned}
\end{equation}

The QS can be seen as an abstract process or quantum protocol, and described in the process matrix formalism \cite{oreshkov2012quantum}.  The process matrix framework describes protocols involving $N$ agents or parties $\{A_i\}_{i=1}^N$. Each agent is associated with a “local laboratory” that has an input quantum system $A_i^I$ and output quantum system $A_i^O$ between which they apply a quantum operation $\mathcal{M}^{A_i}$. The Hilbert space associated to the in and outputs are $\mathcal{H}_{A_i^I}$ and 
$\mathcal{H}_{A_i^O}$, and denoting the space of linear operators on a Hilbert space $\mathcal{H}$ as $\mathcal{L}(\mathcal{H})$, $\mathcal{M}^{A_i}$ consists of quantum channels (completely positive and trace preserving maps) from quantum states in $\mathcal{L}(\mathcal{H}_{A_i^I} )$ to those in $\mathcal{L}(\mathcal{H}_{A_i^O} )$.

The process matrix is a Hermitian operator $W\in\mathcal{L}(\mathcal{H}^{A_1^I}\otimes \mathcal{H}^{A_1^O}\otimes\cdots\otimes\mathcal{H}^{A_N^I}\otimes \mathcal{H}^{A_N^O})$ which describes the environment external to all $N$ labs and models how they can be connected, without assuming a global causal order between the different labs’ operations. It can describe protocols with a fixed order between the agents’ operations: e.g., $A_1$ acts first, then $A_2$, then $A_3$ and so on, as well situations without a definite acyclic order between these agents’ operations.

The process matrix $W_{QS}$ of the quantum switch is a rank-1 projector $W_{QS}=\ketbra{w}{w}_{QS}$, with the process vector:
\begin{equation}
\begin{aligned}
    \ket w_{QS}&=\ket{00}^{C^O_C D^I_C} \kett{\mathbb{1}}^{C^O_T A^I} \kett{\mathbb{1}}^{A^O B^I} \kett{\mathbb{1}}^{B^O D^I_T}\\ &\quad+\ket{11}^{C^O_C D^I_C} \kett{\mathbb{1}}^{C^O_T B^I} \kett{\mathbb{1}}^{B^O A^I} \kett{\mathbb{1}}^{A^O D^I_T},
\end{aligned}
\end{equation}
where $\kett{1}$ is the short-hand notation for a maximally entangled state on the bipartite system. The process vector $\ket{w}_{QS}$ describes a coherent superposition of the order $C\prec A\prec B\prec D$ and $C\prec B\prec A\prec D$, which cannot be written as a probabilistic mixture of causally ordered processes. The subscripts $C$ and $T$ denote control and target. Such a process is called causally non-separable \cite{araujo2015witnessing}, and is referred to as having indefinite causal order. The QS is considered such an example of an indefinite causal order process.

The quantum switch and general indefinite causal order processes can also be understood in terms of a definite cyclic causal structure \cite{Barrett_2021, ferradini2025} (after all, both involve the absence of a definite acyclic causal structure), bearing links to a special subclass of information-theoretic models of closed timelike curves \cite{Araujo_2017}. 

There are several equivalent frameworks for describing such scenarios involving indefinite causal order: as quantum supermaps or higher order transformations \cite{chiribella2013quantum}, process matrices \cite{oreshkov2012quantum}, linear post-selected closed timelike curves \cite{Araujo_2014}, linear mutli-time states \cite{Silva_2017}. As physicality questions for the quantum switch, and related notions like causal non-separability are commonly formulated in the process matrix framework, we refer to this in the following, though the statements can equivalently be applied in the other formalisms.

\section{The quantum switch: connecting (thought) experimental realisations to theoretical descriptions}
\label{appendix: analysis}

  \begin{figure*}[ht!]
        \centering
        \subfloat[Fine-grained description of QS $W_{QS}^f$. Figure taken from \cite{VilasiniRennerPRA, VilasiniRennerPRL}.]{
        \includegraphics[scale=0.5]{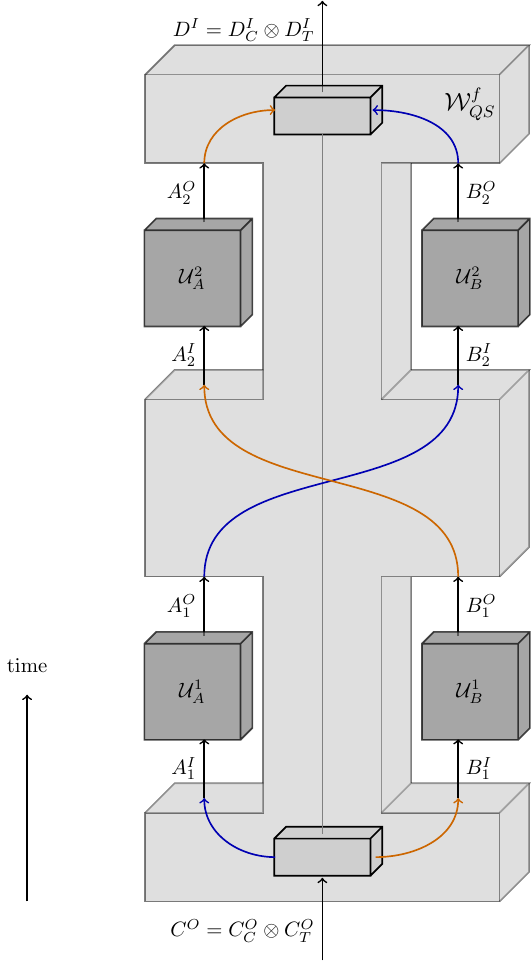}}\qquad\qquad\qquad\subfloat[Routed quantum switch, figure from \cite{Ormrod_2023}.]{\includegraphics[scale=0.35]{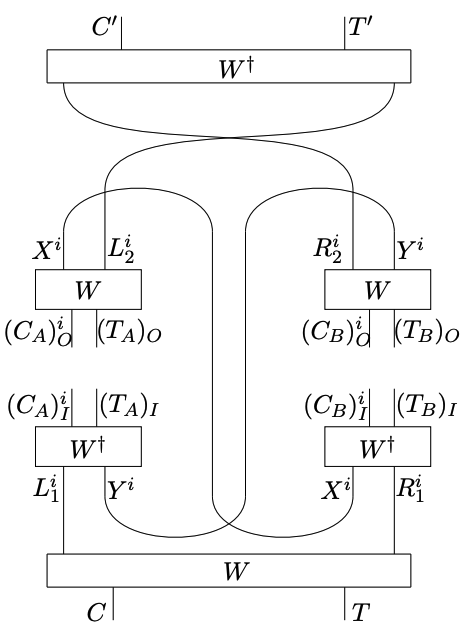}}
        \caption{(a) $W_{QS}^f$ is a quantum process with a definite and acyclic causal order between four operations $\mathcal{U}_{A/B}^{1/2}$ (two for each agent, Alice and Bob) which act on the one-photon (original ``target'') and zero-photon (vacuum) modes. This is a regular time-ordered quantum circuit on vacuum and non-vacuum states where the information-theoretic causal structure of the circuit coincides with the spatio-temporal causal structure of the spacetime diagram in \cref{fig:examples_QS}(a). The two internal boxes which connect the past input to $\mathcal{U}_{A/B}^{1}$ and $\mathcal{U}_{A/B}^{2}$ to the future output, are directly related to the transformation $W_{sup}$ of \cref{eq:Wsup} and its inverse on the 1-message space. The extra gray wire in between additionally encodes the control as an internal memory to allow the operation to be well-defined on the full space of two $d+1$-dimensional systems (see \cite{Portmann2017, VilasiniRennerPRA}) but is not strictly necessary. Similar models for specific types of QS experiments were initially proposed in \cite{Portmann2017,Paunkovic2020}, which were generalised and formalised as fine-graining of the process matrix in \cite{VilasiniRennerPRA, VilasiniRennerPRL}. (b) This is described by a so-called routed supermap, an object that acts on a restricted set of allowed operations (routed unitaries) in the slots. There is are two slots, for Alice and Bob's operations respectively, each acts on a control and a (non-vacuum) target, while constrained to act coherently on the control (\cref{eq: Alice_operation_routed}). The superscripts $i$ indicate such \emph{sectorial constraints} (see \cite{Ormrod_2023}). The operations $W$ here correspond to the operation $W_{sup}$ of \cref{eq:Wsup}. Notice a crucial distinction: Alice and Bob's operations in (a) act on non-vacuum and vacuum, while in (b) they only act on the non-vacuum (control and target). }
        \label{fig:QS_fine}
    \end{figure*}

In the main text (\cref{fig:examples_QS}), we have discussed three (thought) experimental realisations of the quantum switch (QS) which are based on the optical switch (QS$_O$) and the gravitational switch (QS$_G$) that have been widely studied in the literature. A natural question is, what are the accurate theoretical descriptions of these different realisations? The process matrix of the quantum switch, discussed in \cref{QS review}, corresponds to an abstract higher-order quantum transformation that describes the action of QS on the two unitaries $U_A$ and $U_B$. Does this process matrix provide an accurate description of all the realisations of QS, irrespective of the regime (e.g., Minkowski spacetime, quantum gravitational spacetime) of the concerned realisation? Are there any differences between these distinct physical realisations in terms of the events, their localisation or causal structure? These have remained long debated open questions, on which consensus has been lacking. 

Specifically, there have been alternative theoretical descriptions (all based on operational, quantum information-theoretic approaches) proposed for some of these realisations of QS, which are distinct from the original process matrix. Broadly, we distinguish three categories of such descriptions, summarised below (see also \cref{table:appendix}).

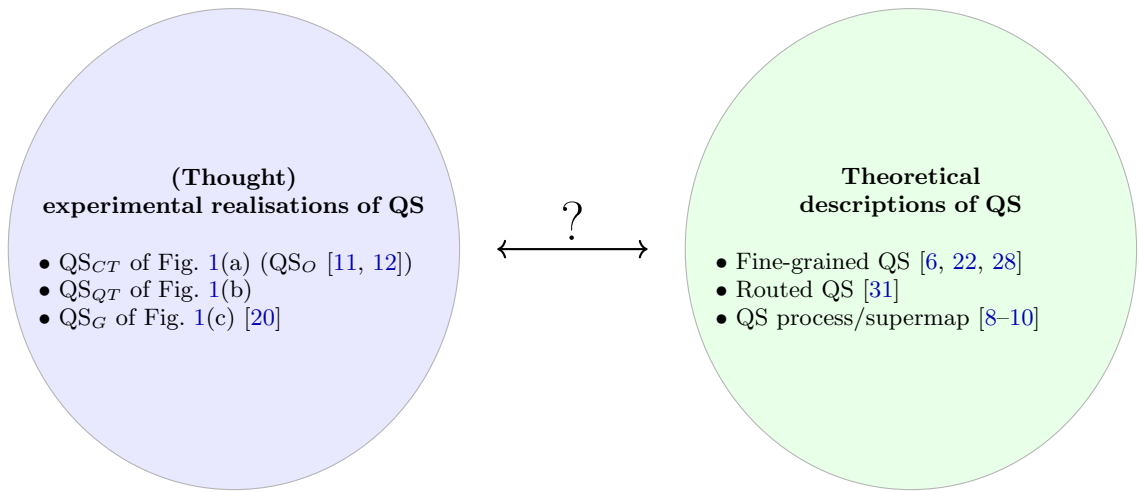
\begin{figure*}[ht!]
\centering
\begin{tikzpicture}
  \draw[fill=blue!30, opacity=0.3] (-4,0) ellipse (3.0cm and 3.2cm);
  \node[align=left, text width=5.2cm] at (-4,0) {
    \centering\textbf{(Thought)\\experimental realisations of QS}\\[1.2em]
    \raggedright
    $\bullet$ QS$_{CT}$ of \cref{fig:examples_QS}(a) (QS$_O$ \cite{Araujo_2014, Procopio_2015})\\ 
    $\bullet$ QS$_{QT}$ of \cref{fig:examples_QS}(b)\\
    $\bullet$ QS$_G$ of \cref{fig:examples_QS}(c) \cite{zych2019bell}
  };

  \draw[fill=green!30, opacity=0.3] (5,0) ellipse (3.0cm and 3.2cm);
  \node[align=left, text width=5.2cm] at (5,0) {
    \centering\textbf{Theoretical\\descriptions of QS}\\[1.2em]
    \raggedright
    $\bullet$ Fine-grained QS \cite{VilasiniRennerPRA, Portmann2017,Paunkovic2020}\\
    $\bullet$ Routed QS \cite{Ormrod_2023}\\
    $\bullet$ QS process/supermap \cite{chiribella2013quantum, oreshkov2012quantum, araujo2015witnessing}
  };

  \draw[<->, thick] (-0.5,0) -- (1.5,0) node[midway, above] {\huge ?};
\end{tikzpicture}

\caption{Despite much theoretical and experimental progress, consensus is severely lacking on how the theoretical descriptions of QS connect to its (thought) experimental realisations. This is a matter of key importance as the descriptions differ in conclusions on events, localisation and causal structure. Our results in \cref{table:main} and \cref{table:appendix} shed light on this issue concretely, our main point being: the same experiment can be associated with distinct theoretical descriptions depending on the Lab from which it is analysed.}
\label{fig: keyquestion}
\end{figure*}

\begin{itemize}

\item {\bf Original process matrix/supermap (Coarse-grained description) \cite{chiribella2013quantum,araujo2015witnessing}, \cref{fig: QS_PM}} As described in \cref{QS review}.  This description is independent of spacetime information or reference d.o.f., involves 2 events corresponding to the operations $U_A$ and $U_B$ of Alice and Bob and exhibits indefinite causal order between these events.

\item {\bf Fine-grained description \cite{VilasiniRennerPRA, VilasiniRennerPRL}, \cref{fig:QS_fine}(a)} A more detailed model incorporating agents' ability to use spatiotemporal degrees of freedom present in the physical realisation when acting on the target, including performing different operations on the target conditioned on the spacetime references. Here, there is a definite and acyclic causal order between $>2$ operations and can be described as a causally separable process matrix/supermap i.e., one that does not exhibit any indefinite causal order. This was specifically proposed for the optical switch experiments that happen in a classical background spacetime (see also \cite{Portmann2017,Paunkovic2020}).

\item {\bf Routed description \cite{Ormrod_2023}, \cref{fig:QS_fine}(b)} This description allows the agents Alice and Bob to access information on the control qubit of the QS, in addition to the target. The agents operations are not arbitrary, e.g., Alice's operation is of the following coherently controlled form on the control and target accessible to her

\begin{equation}
\label{eq: Alice_operation_routed}
\ket{0}\bra{0}\otimes U_A^1 + \ket{1}\bra{1}\otimes  U_A^2,
\end{equation}

where $U_A^{1/2}$ are arbitrary unitaries on the $d$-dimensional target. In particular, measurements or other changes of the control are not allowed. These constraints (``sectorial constraints'') entail that the operations of these agents are routed unitaries, and the corresponding supermap (or process matrix) is a routed supermap. It has been argued that the routed supermap for QS at most exhibits a ``weak form of indefinite causal order'', in contrast to the more strong sense in which the original supermap exhibits ICO.
\end{itemize}

In both the fine-grained and routed QS descriptions, an important transformation involved is the following, which we call $W_{sup}$. This involves considering the original target state $\ket{\psi}$ (a single message or particle such as a photon encoding a qudit) along with a vacuum state $\ket{\Omega}$ modelling the absence of that qudit. 

\begin{equation}
\label{eq:Wsup}
    W_{sup}: (\alpha \ket{0}+\beta \ket{1})\otimes \ket{\psi} \mapsto \alpha \ket{\psi}\otimes\ket{\Omega}+\beta\ket{\Omega}\otimes\ket{\psi}.
\end{equation}

$W_{sup}$ takes as input a qubit, the control, in a non-trivial superposition state and a qudit in state $\ket{\psi}$, and has two outputs, each of $d+1$ dimensions in general: each output (typically linked to Alice and Bob) either contains the qudit state $\ket{\psi}$ or the 1-dimensional vacuum state $\ket{\Omega}$ depending coherently on the control qubit. When restricting to the subspace of the output having exactly 1 qudit (at one of the outputs and vacuum at the other), $W_{sup}$ is a unitary, and $W_{sup}^\dagger$ thus exists in this case \cite{Ormrod_2023}. More generally, without this restriction, the action of $W_{sup}$ can still be recovered as a non-unitary, but physical (i.e. CPTP) quantum channel \cite{Portmann2017,VilasiniRennerPRA}. See \cref{fig:QS_fine} for how this operation features within the fine-grained and routed descriptions.

The literature has been divided on which of these three is the correct description of the performed optical experiments in Minkowski spacetime QS$_O$, as well as how they link to the gravitational quantum switch QS$_G$. This lies at the core of a longstanding debate on the interpretation of the QS$_O$ experiments and whether these entail indefinite causal order \cite{chiribella2013quantum,Procopio_2015,Portmann2017,Vilasini_mastersthesis, Oreshkov2019,Paunkovic2020,VilasiniRennerPRA,VilasiniRennerPRL,Ormrod_2023,Felce_2022,Kabel:2024lzr}.

Here, we apply our new approach for defining events and localisation to the QS realisations, to facilitate consensus and clarify these issues, and specifically to address the key question (depicted in \cref{fig: keyquestion}) of how the (thought) experimental realisations of \cref{fig:examples_QS} correspond with these different theoretical descriptions.  This analysis highlights how different perspectives and formal descriptions emerge under distinct assumptions on the choice of a Lab (as defined in the main text) from which the experiment is analyzed.

This reveals certain crucial, previously overlooked aspects (parts of which are highlighted in \cref{table:main}) which also offer an answer to the key question depicted in  \cref{fig: keyquestion}, as summarised in \cref{table:appendix}.

\begin{itemize}
    \item {\bf Events and causality are relative to a Lab:}  The accurate theoretical description for a given (thought) experimental realisation of QS (QS$_O$, QS$_G$, or QS$_{QT}$ of \cref{fig:examples_QS}) depends on the choice of Lab from which it is analysed. Consequently, even within the same  experiment, conclusions about events, localisation, and whether there is indefintie causal order between the events, depend on the choice of Lab.

    \item {\bf Distinct forms of non-localisation:} Different Labs in QS realisations also differ in their reference degrees of freedom relative to which they localise operational events. We will see that the operational property of \emph{relative measurability} of the reference distinguishes two different forms of non-localisation. In QS, this is linked to distinctions in the Hilbert space structure (tensor product v.s. direct sum). In particular, relative measurability of the reference also distinguishes the fine-grained v.s. routed descriptions of QS, and consequently, the resulting forms of temporal non-localisation. 

\end{itemize}

We begin with the fine-grained description for the optical switch (specifically QS$_{CT}$ of \cref{fig:examples_QS}(a)), discussing how it can recover an effective description with behavior analogous to the routed switch when restricting interventions, and under further restrictions on the allowed interventions (which allows to disregard the reference d.o.f.) one recovers a behaviour similar to the process matrix model. However, we observe key distinctions in events, localisation and causality at each of these stages of the ``coarse-graining'', which highlights that these descriptions are physically and operationally distinct despite the possibility to mathematically map between them. We then discuss how these apply to the gravitational switch QS$_G$, and the additional realisation QS$_{QT}$ of \cref{fig:examples_QS}(b).

\subsection{The optical switch: fine-grained and coarse-grained descriptions}

The QS$_O$ experiment implements the same overall transformation \eqref{eq: QS_action} from the past (agent $C$) to the future (agent $D$) as the quantum switch process matrix in \cref{fig: QS_PM}. This transformation is also recovered in the fine-grained and routed QS descriptions of \cref{fig:QS_fine}. Let us now see how these three descriptions differ, and how they link to the (thought) experimental realisations of \cref{fig:examples_QS}. We link these three descriptions to three distinct sets of assumptions, $\mathcal{A}_{fine}$, $\mathcal{A}_{effective}$ and $\mathcal{A}_{coarse}$ on the Lab $A$ associated to the agent Alice. In all cases, we take the environment d.o.f. which is external to the Lab, as given by the control qubit of the QS.

\subsubsection*{Fine-grained description which includes the vacuum}
The fine-grained view arises from the following assumption on the Labs. Here we discuss a simplified version of the fine-grained description that only uses a particular type of time reference, similar fine-graining arguments apply generally to any experiment in a classical background spacetime \cite{VilasiniRennerPRA, VilasiniRennerPRL}.
\\

\textbf{Assumption} $\mathcal{A}_{fine}$:
    \begin{itemize} 
        \item {\bf Reference} The Lab of Alice (and similarly Bob) are localised relative to the background spacetime, and the reference d.o.f.\  is given by time as specified by a physical clock. In the simplest version of the fine-grained argument, only two times $t_1,t_2>t_1$ will be relevant, and we can take $\mathcal{P}_A:=\{t_1,t_2\}$.\footnote{The argument can be readily also extended to the case where multiple times are involved, and to cover an interesting class of QS$_O$ experiments with long coherence times where the time of arrival is not perfectly entangled, but only rather weakly entangled with the control. See \cite{VilasiniRennerPRA} for details.} 
        \item {\bf Target} The target d.o.f.\  includes the photon in QS$_O$ (that can be modeled as a qudit in $\mathbb{C}^d$) along with the zero-photon mode (modeled by a vacuum state $\ket{\Omega}$). 
        \item {\bf Set of operations}  The physical operations of interest are defined on the above reference and target d.o.f.\ , and allow for different operations to be in-principle performed on the target controlled on the reference information set $\mathcal{P}_A:=\{t_1,t_2\}$.
    \end{itemize}

These assumptions are involved when we wish to analyse the events and causal structure from the perspective of the physical experimenters, who have access to their clocks in order to make sense of the experiment. The inclusion of the vacuum allows to distinguish the time on the clock vs the time of arrival of the photon, which are distinct physical d.o.f.: if Alice knows that the photon arrives to her at the earlier time $t_1$ in the QS$_O$ experiment, she knows that the control is $\ket{0}$ and the photon will not arrive to her (i.e., she receives the vacuum) at time $t_2$. However, if Alice knows that the clock reads time $t_1$, she knows nothing about the whereabouts of the photon or the control qubit. This is an important distinction to keep in mind, which captures the fact that the time $t$ on the lab's clock is a relatively measurable reference for this Alice (\cref{table:main}).

Now, if Alice and Bob's Labs respect the assumptions $\mathcal{A}_{fine}$, they can in-principle apply different unitaries, $\mathcal{U}_{A/B}^1$ vs $\mathcal{U}_{A/B}^2$ on $T_A$ at $t_1$ and $t_2$ of their clocks. This is physically possible in this experimental set-up e.g., by rotating the waveplate in a time-dependent manner, independently of whether or not the photon arrives at the waveplate, this is related to the property of \emph{independent intervenability} also observed in the double slit experiment in \cref{sec: doubleslit}. To capture this, the above unitaries must act on $T_A$ which includes the vacuum, acting as the identity on this 1-dimensional subspace (if no photon goes in, no photon comes out) and applying some unitaries $U_{A/B}^1$ vs $U_{A/B}^2$ on the non-vacuum subspace of the photon. As the time of arrival of the photon is entangled with the control, the resulting transformation differs from \cref{eq: QS_action} while still preserving the coherence of the superposition:
\begin{equation}
\label{eq: OQS_action}
\begin{aligned}
    &\quad(\alpha \ket0+\beta \ket1)_C \otimes \ket \Psi_T \\ &\mapsto \alpha \ket0_C \otimes U^2_B U^1_A \ket \Psi_T + \beta \ket1_C \otimes U^2_A U^1_B \ket \Psi_T.
\end{aligned}
\end{equation}

In the following, we focus on Alice (the analysis for Bob is identical). The above makes it clear that the two unitaries $\mathcal{U}_A^{1}$ and $\mathcal{U}_A^{2}$ independently act on the vacuum extended target $T_A$ are different times. In the fine-grained description of \cref{fig:QS_fine}, this is captured by associating to Alice two inputs, $A_1^I$ and $A_2^I$ (and similarly two outputs $A_1^O$ and $A_2^O$), corresponding to the two times $t_1$ and $t_2$ respectively at which she acts. Each such input (similarly output) is linked to a Hilbert space  

\begin{equation}
    \label{eq: vacext_space}
    \mathcal{H}_{A^I}^t:=(\mathbb{C}^{d}\oplus \ket{\Omega})\otimes \ket{t},
\end{equation}

where $d$ is the dimension of the target system (typically taken as $d=2$ in the quantum switch process matrix), $\ket{\Omega}$ represents the vacuum state, and $t\in\{t_1,t_2\}$. The operations $\mathcal{U}_{A}^{1}$ acts on these vacuum-extended spaces as follows, capturing the intuition that in the experiment of interest, when no photon enters Alice's operation (the waveplate), no photon exits it.

\begin{align}
\label{eq: vacuum_ext_op}
    \begin{split}
        \mathcal{U}_A^{1}:\ket{\Omega}\ket{t_{1}} &\mapsto \ket{\Omega}\ket{t_{1}},\\
        \mathcal{U}_A^{1}:\ket{\psi}\ket{ t_{1}} &\mapsto (U_A^1\ket{\psi})\ket{ t_{1}},
    \end{split}
\end{align}
and similarly for $\mathcal{U}_A^2$. Here, $\psi\in \mathbb{C}^d$ represents any non-vacuum photon state, and $U_A^1$ is a unitary transformation acting on it.\footnote{For simplicity, we have taken these operations occur instantaneously (i.e., with the same input and output times). However, the original framework models each operation causally, allowing for nonzero durations. This can be accounted for by readily adapting our analysis to that case, or simply regarding each $t$ label here as a time interval within which the in and outputs are (causally) produced.}

Next, consider Alice's fine-grained operations $\mathcal{U}_A^1$ and $\mathcal{U}_A^2$, as well as the overall operation $\mathcal{U}_A$ which is formed by the composition of  $\mathcal{U}_A^1$ and $\mathcal{U}_A^2$. Although $\mathcal{U}_A^1$, $\mathcal{U}_A^2$, and $\mathcal{U}_A$ can all be operations performed by the same physical agent, Alice, it is useful—following standard quantum information conventions—to treat them as associated with three distinct agents, labeled $A_1$, $A_2$, and $A$. This distinction arises because they act on different input and output spaces. Indeed, this perspective is consistent with the process matrix framework, where each agent or lab corresponds to a ``slot'' or ``locus of intervention'' in which they can apply an operation from a set of quantum instruments. In the fine-grained description, there are separate slots for $A_1$ and $A_2$, making it a four-party process. Crucially, the initial conditions $\mathcal{I}_{A_1}$, $\mathcal{I}_{A_2}$ and $\mathcal{I}_A$, associated with $A_1$, $A_2$ and $A$ (as given by the description of \cref{fig:QS_fine}), are distinct because they act on different (sub)spaces of the reference and target.

The overall operation $\mathcal{U}_A$ acts on the tensor product of the (vacuum-extended) Hilbert spaces, \cref{eq: vacext_space} corresponding to the two times,  $\mathcal{H}_{A^I}^{t_1} \otimes \mathcal{H}_{A^I}^{t_2}$,
where the tensor product structure reflects the independent intervenability at these two times (such as the in-principle ability to independently apply any operations, including preparing independent states on these inputs). Notice that the operation $\mathcal{U}_A^1$ of \cref{eq: vacuum_ext_op} can be equivalently seen as a controlled operation (controlled on the second subsystem carrying the time information):

\begin{equation} \label{eq: Alice_finegrained} \mathcal{U}_A^1=\Big( U_A^1\oplus \id_{\Omega}\Big)\otimes \ket{t_1}\bra{t_1}: \mathcal{H}_{A^I}^{t_1}\mapsto \mathcal{H}_{A^O}^{t_1}. \end{equation}

Here, $U_A^1$ is defined on the $d$-dimensional non-vacuum subspace, and $\id_\Omega$ is the identity on the 1-dimensional vacuum subspace (see also \cref{eq: vacext_space}). Indeed, the initial conditions $\mathcal{I}_{A_1}$ for $A_1$'s Lab only deliver states in $\mathcal{H}_{A^I}^{t_1}$ to the input of their operation $\mathcal{U}_{A_1}$, which always have a fixed time $\ket{t_1}$. This fact can be physically confirmed by measuring the lab's clock at the time when this operation is applied, and as noted before, this will not affect or decohere the experiment. Therefore, $\mathcal{U}_A^1$ is a $t_1$-localised event, and similarly, $\mathcal{U}_A^2$ is a $t_2$-localised event for for Labs respecting $\mathcal{A}_{fine}$, where $\mathcal{P}_A:=\{\ket{t_1},\ket{t_2}\}$. On the other hand, $\mathcal{U}_A$ is a non-localised event in this case. More, explicitly, this can be seen as follows.

{\bf Non-localisation at the fine-grained level} 
To make a more explicit connection to the notion of a Lab which includes a time reference as in assumption $\mathcal{A}_{fine}$, we can represent the action of $\mathcal{U}_A$ by sequentially composing $\mathcal{U}_A^1$ and $\mathcal{U}_A^2$, while also including a time-shift operation $T^{+1}:\ket{t_1} \mapsto \ket{t_2}$ acting solely on the reference (time) degree of freedom between them. This makes it clear that measuring the time before and after $\mathcal{U}_A$ yields different results ($t_1$ and $t_2$), highlighting that it is a time-extended (i.e., and hence non-localised) event for the chosen Lab. Importantly, measuring time in this way (e.g., consulting a clock on the wall) does not disturb any aspect of the QS$_O$ experiment, specifically the coherent superposition involved in the quantum switch. Therefore the reference (clock) is relatively measurable in the QS$_O$ experiments and in the fine-grained description.

\subsubsection*{An effective description which ignores the vacuum} 

We can show that, from the fine-grained model which includes the vacuum, one can extract an effective description which ignores the vacuum, under the guarantee that each party acts only on one $d$-dimensional non-vacuum d.o.f., albeit not localised in time. 
The operation $\mathcal{U}_A$ acts on the tensor product space $\mathcal{H}_{A^I}^{t_1} \otimes \mathcal{H}_{A^I}^{t_2}$. However, in the standard operation of the quantum switch (where only one photon is present in the experiment), the inputs to $\mathcal{U}_A$ lie in the subspace  

\begin{equation}
    \label{eq: subspace_tensorprod}
    \text{Span}\Big(\{\ket{i,t_1} \otimes \ket{\Omega, t_2},\ket{\Omega,t_1} \otimes \ket{j,t_2}\}_{i,j\in\{0,...,d-1\}}\Big).
\end{equation}

This corresponds to a single non-vacuum $d$-dimensional system being present at $t_1$ or $t_2$ or a superposition thereof, with the vacuum at the other time representing the absence of said system.
 This subspace is isomorphic to the direct sum $(\mathbb{C}^d\otimes \ket{t_1})\oplus (\mathbb{C}^d\otimes \ket{t_2})$ of two non-vacuum $d$-dimensional Hilbert spaces linked to the two times, using the isomorphism $\ket{i,t_1}\otimes\ket{\Omega, t_2}\equiv \ket{i,t_1}$, $\ket{\Omega,t_1}\otimes \ket{j,t_2}\equiv \ket{j,t_2}$. Thus the inputs to Alice's operation effectively lie in 

\begin{equation}
    \label{eq: subspace_directsum}
    \text{Span}\Big(\{\ket{i,t_1},\ket{j,t_2}\}_{i,j\in\{0,...,d-1\}}\Big).
\end{equation}

We can now define $\mathcal{U}_A^{eff}$ to be the effective action of Alice's operation on this subspace, as 

\begin{equation}
\label{eq: Alice_operation_coarse}
    \mathcal{U}_A^{eff} = U_A^1 \otimes \ket{t_1}\bra{t_1} + U_A^2 \otimes \ket{t_2}\bra{t_2}.
\end{equation}

{\bf Non-localisation at the effective level.}  When the control is initially in a superposition state (as is the case in QS) this operation acts on states with non-zero amplitudes for the two times, therefore, measuring the time will not yield a deterministic outcome and $\mathcal{U}_A^{eff}$  is a non-localised event relative to the chosen Lab which includes a time reference (and initial conditions). Unlike in the fine-grained description, however, measuring time in this case collapses the superposition of the quantum switch. In other words, the time reference involved in this effective description is not relatively measurable.

\subsubsection*{Comparing the effective description to the routed and coarse-grained models}

Notice that up to the relabelling $\ket{t_1}\mapsto \ket{0}$ and $\ket{t_2}\mapsto \ket{1}$, and ordering of the two systems, Alice's operation in the effective description (\cref{eq: Alice_operation_coarse}) coincides with that of the routed QS description (\cref{eq: Alice_operation_routed}). However, this alone is insufficient to conclude that we uniquely recover the routed QS description of \cref{fig:QS_fine}(b) for the entire protocol.

To see this, consider the process matrix of the quantum switch \cref{fig: QS_PM}, and prepare two additional qubits $R_A$ and $R_B$ in the past (with orthonormal basis states $\ket{t_1}$, $\ket{t_2})$) which are entangled with the control, thus replacing the initial state $\alpha\ket{0}+\beta\ket{1}$ of the control with
\begin{equation}
    \label{eq: ref_entanglement}    
\alpha\ket{0}_C\ket{t_1}_{R_A}\ket{t_1}_{R_B}+\beta\ket{1}_C\ket{t_2}_{R_A}\ket{t_2}_{R_B}
\end{equation}

The additional systems $R_A$ and $R_B$ serve as ancillas playing the role of a reference degree of freedom for time, and they are sent to Alice and Bob respectively, along with the usual $d$-dimensional target system on which each of them acts. 

If we now apply the effective (coarse-grained) operations defined in \cref{eq: Alice_operation_coarse} for Alice and Bob, we recover the same transformation as \cref{eq: OQS_action}, but with $\ket{0}_C$ in the control replaced by $\ket{0}_C\ket{t_1}_{R_A}\ket{t_1}_{R_B}$ and similarly for $\ket{1}_C$. 

In the routed QS description, the (sectorised) controls $C_A$ and $C_B$ of the two agents plays the role of the reference ancillas $R_A$ and $R_B$. However, these no longer appear in the global future as in the above modification of the original process matrix, the routed QS exactly recovers the transformation \cref{eq: OQS_action}. 
Independently of this distinction, both the routed description and the above mentioned modification of the process matrix respect the following assumption on the Labs.
\\ \\

\textbf{Assumption} $\mathcal{A}_{effective}$:
\begin{itemize}
    \item {\bf Reference} The Lab of Alice (and similarly Bob) use as reference a quantum d.o.f.\  with a 2-element set $\mathcal{P}_A:=\{t_1,t_2\}$ that is entangled with the control qubit of the quantum switch (in the basis $\{\ket{t_1}, \ket{t_2}\}$).
     \item {\bf Target} $T_A$ only includes the non-vacuum description of the target qudit in the QS experiment (photon, in the QS$_O$). 
      \item {\bf Set of operations} The physical operations of interest act on the above mentioned reference and target, and allow different operations on the target, coherently controlled on the reference basis  $\{\ket{t_1}, \ket{t_2}\}$.
\end{itemize}

Notice that any Lab for a QS experiment which satisfies $\mathcal{A}_{effective}$ cannot have a measurable reference because the reference-control entanglement would imply that $\mathcal{P}_A$-level (here given by the basis $\{\ket{t_1},\ket{t_2}\}$) measurements on the reference would decohere the superposition of orders in the QS. 

Furthermore, the effective description derived from the fine-grain model also respects $\mathcal{A}_{effective}$, and we thus lose the property of relative measurability under this mapping.
This highlights a key physical distinctions between the fine-grained vs the effective as well as the related routed QS descriptions, despite mathematical similarities between them for situations of interest. Crucially, $t_1$ and $t_2$ in $\mathcal{A}_{fine}$ correspond to the times as given by a clock but in $\mathcal{A}_{effective}$, to the time of arrival of the non-vacuum target (photon) at Alice's lab, which are distinct physical d.o.f.

Finally, we also observe that in the case that the operations on the (non-vacuum) target are the same independently of the reference state, we obtain again the transformation of the original QS process matrix \cref{eq: QS_action}. The assumptions underlying the original process matrix description can be captured as follows.
\\ \\
\textbf{Assumption} $\mathcal{A}_{coarse}$:
\begin{itemize}
    \item {\bf Reference} The reference information set of the Lab of Alice (and Bob) has a single element.
    \item {\bf Target} $T_A$ only includes the non-vacuum description of the target qudit (photon, in the QS$_O$).
    \item {\bf Set of operations} Arbitrary operations on the target are allowed, but there are no non-trivial operations on the reference.
\end{itemize}

The single element of the reference can be seen as simply labeling the target $T_A$ qudit on which the lab acts. In the process matrix description, this is given by the in and output systems $A^I$ and $A^O$ of Alice's lab which are both qudits. Our definitions then imply that all operations $O_A\in\mathcal{O}_A$ of the Lab are localised at the single reference element labelled ``$T_A$'', i.e., they are $T_A$-localised in the subsystem localisation sense\footnote{See \cref{sec: definitions} for how subsystem information can be encoded in the reference and how subsystem localisation can be recovered in our framework. }.

\begin{table*}[ht!]
\centering
\resizebox{\textwidth}{!}{
\begin{tabular}{|c|c|c|c|c|c|c|c|}
    \hline
    Theoretical descriptions & Non-trivial reference? & Agent acts on vacuum? & Rel. measurability? & ICO? & QS$_{CT}$ & QS$_{QT}$ & QS$_{G}$ \\ 
    \hline 
    Fine-grained & Yes & Yes & Yes & No & Alice’s Lab with $(x,t)$ & Claire’s Lab with $(x,t)$ & Open q: follow-up work \\
    Effective & Yes & No & No & Weak ICO?& Alice's Lab with $t_{arr}$ & Alice’s Lab using $(x,t)$, $a$ & Alice’s Lab using $a$ \\
    Coarse-grained & No & No & Yes & Yes & Any Lab with triv. reference & Alice’s Lab using $\tau$ & Alice’s Lab using $\tau$ \\
    \hline
\end{tabular}
}
\caption{Summary of the analysis in \cref{appendix: analysis}: Motivated by the key question in \cref{fig: keyquestion}, we examined three distinct theoretical descriptions of the quantum switch, fine-grained, routed, and coarse-grained, associated with three corresponding sets of assumptions, $\mathcal{A}_{fine}$, $\mathcal{A}_{effective}$ and $\mathcal{A}_{coarse}$, and linked these with the QS (thought) experiments QS$_{CT}$, QS$_{QT}$ and QS$_G$ of \cref{fig:examples_QS}. This shows that the same physical experiment can have distinct theoretical descriptions depending on the choice of Lab from which it is analysed (incorporating choices of criteria (a)-(c) of \cref{Introduction}), which lead to different conclusions on events and causality. The routed QS description of \cite{Ormrod_2023}, previously argued to display only a weak form of indefinite causal order (ICO), satisfies the assumptions $\mathcal{A}_{effective}$ identified here, but not $\mathcal{A}_{fine}$ or $\mathcal{A}_{coarse}$, we thus place it in the “effective” category, though we noted that it is not uniquely determined by those assumptions. Here, $x$, $t$ denote background space and time coordinates given by physical references e.g., clocks, $t_{arr}$ denotes the time of arrival (of the photon in QS$_{CT}$), $\tau$ is proper time and $a$ denotes acceleration (in QS$_{QT}$, QS$_G$). Finally, we note that the case of Alice's Lab using proper time $\tau$ as a reference in QS$_{QT}$ and QS$_G$ is analogous to the case with $|\mathcal{P}_A|=1$, since only a single value $\tau_*$ (or a unique ``location'' relative to that reference) is operationally relevant per agent, regardless of their order of action. }
\label{table:appendix}
\end{table*}

{\bf Time non-localisation and conclusions for QS$_O$} We have analysed three distinct theoretical descriptions of the quantum switch, based on different assumptions ($\mathcal{A}_{fine}$, $\mathcal{A}_{effective}$, and $\mathcal{A}_{coarse}$). Notably, only the fine-grained and effective descriptions incorporate a non-trivial reference degree of freedom for Alice and Bob’s lab, which can be associated some notion of time involved the optical quantum switch (QS$_O$) experiments. Both involve non-localisation of the agents’ operation in “time” (consistent with previous works \cite{Oreshkov2019,zych2019bell,Giacomini_2019,VilasiniRennerPRA,VilasiniRennerPRL, kabel2024}).

However, our analysis demonstrates that these are physically distinct forms of time non-localisation. In the fine-grained case, time corresponds to a relatively measurable reference (e.g., given by a clock on the wall, $t$), whereas in the effective description it corresponds to a relatively non-measurable quantum reference (e.g., the photon’s time of arrival, $t_{arr}$). 

This distinction has operational significance: in real world QS$_O$ experiments, if the labs have access to classical clocks that reflect the time coordinate, measurability with respect to this reference can be empirically verified without disturbing the QS operation. As our analysis shows, only the fine-grained description \cite{VilasiniRennerPRA} reproduces this feature of the QS$_O$ experiments, unlike the coarse-grained (original process matrix/supermap, \cref{QS review}) description  or the routed QS description \cite{Ormrod_2023}, even if these may emerge as an effective model from the fine-grained description. The fine-grained model is therefore the appropriate one for these experiments when analysed from classical Labs that use their clocks as references. The other models remain relevant for other experiments (e.g., QS$_{QT}$, QS$_G$) and choices of references (e.g., time of arrival, acceleration). These differences yield distinct conclusions regarding the presence of indefinite causal order, as summarised in \cref{table:appendix}.

More generally, we note that explicit fine-grained models having a definite and acyclic causal order (analogous to \cref{fig:QS_fine} exists for a large class of process matrices with indefinite causal order \cite{Salzger2024}. Effective descriptions and mappings between the fine and coarse grained descriptions of such processes are studied in \cite{Vilasini_QPL2020,Salzger2024,SalzgerThesis}, which also provides an approach to address the question of which processes can be physically realised in a classical spacetime (to the same extent that QS was realised in optical experiments).

\subsection{Gravitational quantum switch (and QS$_{QT}$)}

In both the Gravitational Quantum Switch (QS$_G$) and QS$_{QT}$ described in the main text, the reference degrees of freedom $R_A$ for Alice (and similarly $R_B$ for Bob)—such as acceleration in QS$_G$ or spatial distance in QS$_{QT}$—are entangled with the control of the quantum switch. Hence the Labs including these references respects the assumptions $\mathcal{A}_{effective}$ and consequently the routed QS of \cref{fig:QS_fine}(b), provides a more accurate description of these scenarios compared to the fine-grained model of \cref{fig:QS_fine}(a) or the original QS process matrix of \cref{fig: QS_PM}. The argument for the non-localisation of Alice’s and Bob’s operations relative to these references follows identically from the analysis for what we called the ``effective description'' in the previous section.

A fine-grained analysis could still be formulated for the Lab of Alice in QS$_G$ using the aforementioned isomorphism embedding a direct sum of spaces into a tensor product of spaces, but it would lack the same physical and operational meaning that the case of QS$_{CT}$ (or equivalently QS$_O$ analysed using Labs respecting $\mathcal{A}_{fine}$). This is precisely because of the distinctions between the Alices in QS$_{CT}$, QS$_{QT}$ and QS$_G$ with regards to relative measurability of their reference d.o.f and ability to independently intervene using the references of their Labs, discussed in \cref{table:main}, and in \cref{sec: doubleslit}, \cref{sec: QSanalysis}. For instance, in contrast to QS$_{CT}$ where the clock is relatively measurable and allows independent interventions for the Alice therein, the acceleration in QS$_G$ or the relative distance in QS$_{QT}$ are not relatively measurable for the respective Alices and do not permit independent interventions to be performed at the two ``locations'' given by that reference. Whether a physically meaningful fine-grained description can exist for the external, far away agent Claire$_G$ in QS$_G$ (as it does for Alice in QS$_{CT}$) is an open question which we address in a follow-up work (as described in \cref{sec: outlook}.

Our analysis indicates that better understanding how events, localisation and causality depend on the choice of Lab, can help clarify relations between different causality frameworks used to model quantum switch like processes, such as causal boxes \cite{Portmann2017,Vilasini_QPL2020,Salzger2024,SalzgerThesis}, routed quantum circuits \cite{Vanrietvelde_2021,Ormrod_2023}, time-delocalised subsystems \cite{Oreshkov2019}, quantum circuits with quantum control of causal order \cite{Wechs_2021}, and quantum reference frame-based formalisms (e.g., \cite{Giacomini_2019,kabel2024}).  Moreover, we have seen that relative measurability of the reference d.o.f.\  and the ability of an agent to independently intervene relative to that reference (which has a bearing on the Hilbert space structure) are closely related in the QS examples. Further work is needed to formalise this property of independent intervenability in general and link it to relative measurability. This could shed further light on classical vs non-classical features of the reference d.o.f.\  and corresponding localisation of events.
We hope that the proposal intiated in this work
contributes to broader investigations on these aspects,
also beyond the QS scenarios.

\end{document}